\documentclass[twocolumn,preprintnumbers,superscriptaddress,nofootinbib,aps,prd,floatfix]{revtex4}

\usepackage{graphicx}
\usepackage{amsmath,mathenv,bbm,amsfonts,color}
\usepackage{subfigure}
\usepackage{graphicx}
\usepackage{cases}
\usepackage{enumerate}
\usepackage{enumitem}
\usepackage{slashed}

\newenvironment{cedescription}{%
   
   \begin{description}[leftmargin=0.2cm, style=sameline]%
}{%
   \end{description}%
}

\preprint{IPPP/15/05} 
\preprint{DCPT/15/10}

\begin{document}

\title{On-shell interference effects in Higgs final states}

\begin{abstract}
  Top quark loops in Higgs production via gluon fusion at large
  invariant final state masses can induce important interference
  effects in searches for additional Higgs bosons as predicted in,
  {\it e.g.}, Higgs portal scenarios and the MSSM when the heavy
  scalar is broad or the final state resolution is poor. Currently,
  the limit setting as performed by both ATLAS and CMS is based on
  injecting a heavy Higgs-like signal neglecting interference
  effects. In this paper, we perform a study of such ``on-shell''
  interference effects in $pp\to ZZ$ and find that they lead to a
  $\lesssim{\cal{O}}(30\%)$ width scheme-dependent modification of the
  signal strength. Including the continuum contributions to obtain
  {\it e.g.}  the full $pp\to ZZ \to 4\ell$ final state, this
  modification is reduced to the 10\% level in the considered
  intermediate mass range.
\end{abstract}

\author{Christoph Englert} \email{christoph.englert@glasgow.ac.uk}
\affiliation{SUPA, School of Physics and Astronomy,\\University of
  Glasgow, Glasgow G12 8QQ, United Kingdom\\[0.1cm]}

\author{Ian Low} \email{ilow@northwestern.edu}
\affiliation{High Energy Physics Division, Argonne National
  Laboratory, Argonne, IL 60439, USA\\[0.1cm]}
\affiliation{Department of Physics and Astronomy, Northwestern
  University, Evanston, IL 60208, USA\\[0.1cm]}

\author{Michael Spannowsky}\email{michael.spannowsky@durham.ac.uk} 
\affiliation{Institute for
  Particle Physics Phenomenology, Department of Physics,\\Durham
  University, Durham DH1 3LE, United Kingdom\\[0.1cm]}

\maketitle

%%%%%%%%%%%%%%%%%%%%%%%%%%%%%%%%%%%%%%%%%%%%%%%%%%
\section{Introduction}
\label{sec:intro}
The discovery of the Higgs boson~\cite{hatlas,hcms,hzz} with signal
strengths in good agreement with the Standard Model (SM) expectation
marks the end of the endeavor to complete the SM particle
spectrum. The Higgs mechanism, {\it i.e.} the non-linear realization
of gauge invariance with a non-trivial vacuum configuration is the
only known theoretically consistent QFT framework that allows to
include gauge boson masses in non-abelian field theories. Furthermore,
as formulated in the minimal set-up of the SM, fermion masses can be
included through non-trivial and chirality-breaking interactions with
this vacuum.

While the semi-classical limit as expressed in the tree-level
Lagrangian captures all these effects at face value, the implications
beyond leading order are less obvious. Unitarity, or equivalently
electroweak renormalizability, shapes the phenomenology of the
physical Higgs boson by directly linking the fermion and gauge boson
sectors~\cite{Chanowitz:1978mv}. Hence, modifying the couplings of the
Higgs to fermions or gauge bosons in a non-consistent way typically
introduces theoretical shortcomings, which can be resolved by
understanding the SM as a low-energy effective field theory
(EFT)~\cite{dim6,dim6e,dim6r,dim6g,dim6gg,dim6f,john,trott,espinosa,Englert:2014cva,Ghosh:2014wxa}.

In non-EFT extensions of the SM, the currently allowed range of Higgs
couplings can be mapped onto a prediction of additional resonances
that contribute to the restoration of high scale unitarity through
compensating a deviation of the observed Higgs couplings from the
SM. A minimal framework that has been adopted by the experiments to
look for such states is the so-called Higgs portal
scenario~\cite{portal_orig}, which provides a well-defined setting to
model and interpret searches for additional SM-like Higgs
resonances~\cite{portal}, and, at the same time, interfaces the SM
with known BSM effects~\cite{choiportal,portal2,portal3,portal4}.

One of the most promising processes to search for such an additional
heavy state is Higgs production via gluon fusion with subsequent decay
to leptons $pp\to ZZ\to 4\ell$ (a first complete analysis was
presented in \cite{bsm2}) or semileptonic $ZZ$
decays~\cite{Hackstein:2010wk}, depending on the mass of the heavy
Higgs-like state. The $pp\to ZZ$ channels have gained a lot of
interest recently in the context of ``off-shell'' Higgs
measurements~\cite{Kauer:2013qba,melnikov,ciaran,cmswidth,atlaswidth}
(see also~\cite{Dixon:2013haa2}), in particular as probe of new
physics~\cite{bsm,bsm2,fermilab1,fermilab2,bsmh,mit,tilman}. Due to an
a priori large light Higgs contribution at large invariant final state
masses \cite{Kauer:2013qba}, setting limits by injecting a signal
hypothesis without including interference effects can in principle
lead to a quantitatively wrong exclusion in the absence of an excess.

In Monte Carlo programs that underpin this limit setting procedure, we
typically employ a Breit-Wigner propagator
\begin{equation}
  \label{eq:breitw}
  \Delta_h (p^2,m_h^2,\Gamma_h)= {i \over p^2-m_h^2+im_h\Gamma_h}
\end{equation}
to ensure a correct behavior at low Higgs boson virtualities (this
means in particular a non-diverging cross section). However, the
Breit-Wigner distribution cannot be motivated from first-principle
Quantum Field Theory and typically is tantamount to unitarity
violation \cite{Stuart:1991xk,Hwidth,Hwidth2}.

That said, the structure of Eq.~\eqref{eq:breitw} is reminiscent of a
Dyson-resummation of the imaginary part of the Higgs self-energy
$\Sigma_H(p^2)$ for time-like momenta, which is related to its total
decay width via
\begin{equation}
  \label{eq:impart}
  \hbox{Im}\{\Sigma_H(p^2)\} \sim {p^2\Gamma_h\over m_h}\,.
\end{equation}
It should be stressed that this relation can only serve as a scaling
argument for the Higgs boson, for details see {\it
  e.g.}~\cite{Hwidth2}. In any case, the Breit Wigner distribution, especially for
space-like momenta, is an ad-hoc substitution
$\Gamma_h\to \Gamma_h m_h^2/p^2$.

%%%%%%%%%%%%%%%%%%%%%%%%%%%%%%%%%%%%%%%%%%%%%%%%%%
\begin{figure}[!t]
  \begin{center}
    \includegraphics[width=0.46\textwidth]{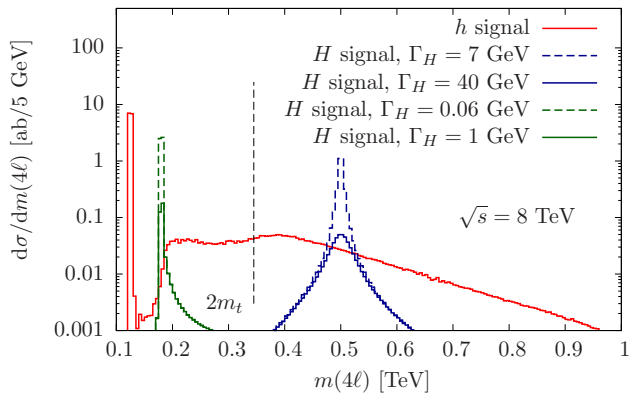}
    \caption{\label{fig:mzz} The distributions are obtained with a
      naive Breit-Wigner propagator.}
    \end{center}
\end{figure}
%%%%%%%%%%%%%%%%%%%%%%%%%%%%%%%%%%%%%%%%%%%%%%%%%%

A consistent transition to complex mass poles as indicated in
Eq.~\eqref{eq:impart} avoids the theoretical
shortcomings~\cite{Passarino:2010qk}, and unless we do not
artificially split a full scattering amplitude into ``signal'' and
``background'' contributions, there are no ambiguities: The
renormalized scattering amplitude will be gauge-invariant and
unitarity is conserved as a consequence.\footnote{Practical schemes
  such as the complex mass
  scheme~\cite{Denner:2005fg,Denner:2005es,calc} share this property.}
A proper treatment of heavy Higgs signals in $pp\to ZZ$ scattering has
been performed in Ref.~\cite{Hwidth2} in the context of the Standard
Model. 

In the Higgs portal scenario, $pp\to ZZ$ receives an additional
``background'' contribution from the off-shell SM Higgs, which can be
similar in size, Fig.~\ref{fig:mzz}. It is the purpose of this note to
also give a discussion of how important these effects are. In
the spirit of Ref.~\cite{Hwidth2}, the phenomenological difference of
Breit Wigner propagators vs. a theoretically clean definition of
signal strengths from complex poles should be included in
experimental analyses as an additional source of theoretical
uncertainty at the leading-order accuracy that we consider in this
work.

This note is organized as follows: We first quickly review the Higgs
portal scenario in Sec.~\ref{sec:Hport} to make this work
self-consistent before we discuss the light Higgs signal-heavy Higgs
signal interference in
Sec.~\ref{sec:hhinterf}. Section~\ref{sec:allinf} is devoted to a
discussion of the complete heavy Higgs signal-continuum interference.

\section{The setup}
\subsection{The Higgs Portal Scenario}
\label{sec:Hport}
The Higgs portal scenario as introduced in Ref.~\cite{portal_orig} is
an extension of the Higgs sector by another scalar field $\phi$
\begin{multline} 
  \label{eq:portal}
  {\cal{V}}_{\text{Higgs}}=\mu^2_\Phi |\Phi|^2 + \lambda_\Phi |\Phi|^4
  +\tilde\mu^2_\phi |\phi|^2 + \tilde \lambda_\phi |\phi|^4 \\+\eta |\Phi|^2
  |\phi|^2  \,,
\end{multline}
where $\Phi$ denotes the SM Higgs doublet and $\phi$ transforms as a singlet under the SM gauge
interactions. Minimizing the potential for non-trivial $\lambda_\phi$,
we can rewrite Eq.~\eqref{eq:portal} in the standard form 
\begin{alignat}{4}
  \phi=&(v_\phi+\tilde \phi)/\sqrt{2}\,,\\
  \Phi=&(v_\Phi+\tilde \Phi) /\sqrt{2}\,,
\end{alignat}
where $v_{\Phi,\phi}$ are the vacuum expectation values of the
corresponding fields, which are functions of the underlying parameters
in the Lagrangian (for details see Ref.~\cite{portal}). 

The modifications compared to SM Higgs phenomenology are introduced by
a linear mixing between the $\Phi,\phi$ fields that can be
diagonalized with a single orthogonal transformation that relates the
Lagrangian basis $\{\Phi,\phi\}$ to the mass basis $\{h,H\}$,
\begin{equation}
  \label{eq:rot}
  \left(\begin{matrix} \tilde \Phi \\ \tilde \phi \end{matrix} \right)_{\cal{L}} = 
  \left(\begin{matrix} \cos\chi   & -\sin\chi \\ \sin\chi & \cos\chi  \end{matrix} \right)
  \left(\begin{matrix} h \\ H \end{matrix} \right)_{\cal{M}}\,.
\end{equation}
Equation~\eqref{eq:rot} makes apparent that the bulk of the model's
single Higgs phenomenology can be traced back to a single mixing angle,
which universally rescales all Higgs couplings. Although parameter
choices are possible for which the observed 125 GeV boson is the
heavier of the two states, we do not consider this option in the
following (for a recent discussion including electroweak precision
effects see Ref.~\cite{david}).

In its simplest implementation with only one hidden sector field, the
cascade width $H\to hh$ and, hence, the total decay widths are fixed
by the SM sector and the extended symmetry breaking potential and
provide crucial information to reconstruct the model's parameters in
its simplest realization \cite{portal,choiportal}. 

To capture the importance of the on-shell interference, however, we
choose a different approach to include the particle widths in our
simulation by choosing the width of the heavy state as a free
parameter. On the one hand this allows us to scan the impact of the
Higgs width on the mentioned interference effects directly. On the
other, once we allow for the presence of a hidden sector in the
fashion of Eq.~\eqref{eq:portal}, there is no a priori reason why the
boson widths are fixed to their SM-like values times the
characteristic mixing angle supplemented by $H\to hh$. In fact,
allowing for more than a single singlet extension as predicted in many
UV complete scenarios \cite{Jaeckelpaper,choiportal} loosens the tight
correlation of the Higgs phenomenology of Eq.~\eqref{eq:rot} with the
fundamental parameters in the Lagrangian \cite{portal}. While we can
still interpret the Higgs phenomenology in terms of an (effective)
mixing angle due to decreased couplings compared to the SM in this
case, the states' widths become less constrained. From this
perspective, injecting a heavy Higgs signal whilst keeping its width
as a free parameter as performed in recent analyses by the CMS
collaboration \cite{Khachatryan:2014wca} is sensitive to a wider class
of scenarios and provides a phenomenological bottom up approach to
formulate constraints on the presence of extra heavy scalar
resonances. The question of the impact of interference effects, which
is typically neglected in the limit setting procedure, remains as a
crucial systematic uncertainty.

\section{Width and Propagator}
The characteristic structure of Fig.~\ref{fig:mzz} implies a shift of
the $H$ pole in comparison to the on-shell mass when inferred from an
invariant mass measurement. The quantitative effects have been
discussed in Refs.~\cite{bsm2,Maina:2015ela,kauernew} in detail. In
this work we also analyze the impact of the implementation of
propagator on this particular feature.

%%%%%%%%%%%%%%%%%%%%%%%%%%%%%%%%%%%%%%%%%%%%%%%%%%
\begin{figure*}[!t]
  \begin{center}
    \includegraphics[width=0.43\textwidth]{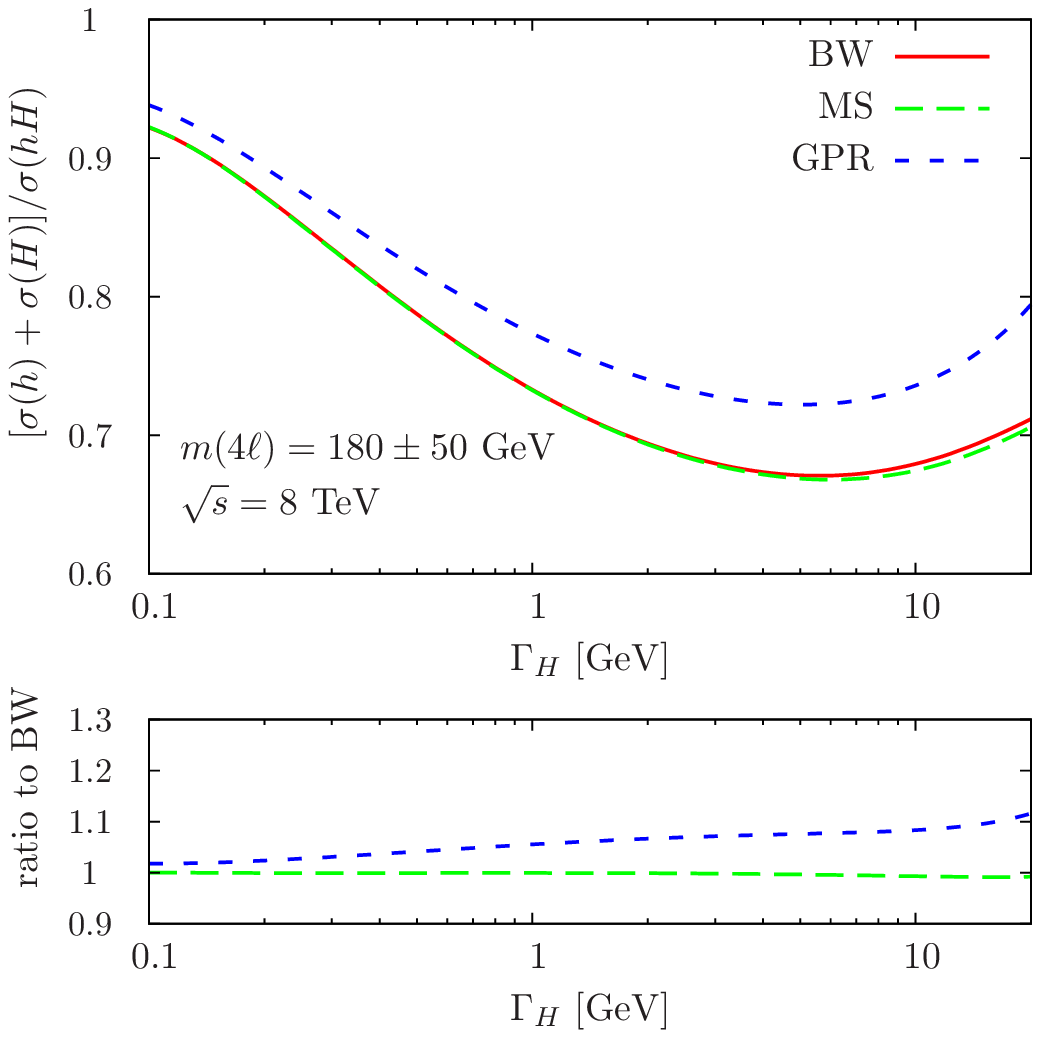}\hspace{1cm}
    \includegraphics[width=0.43\textwidth]{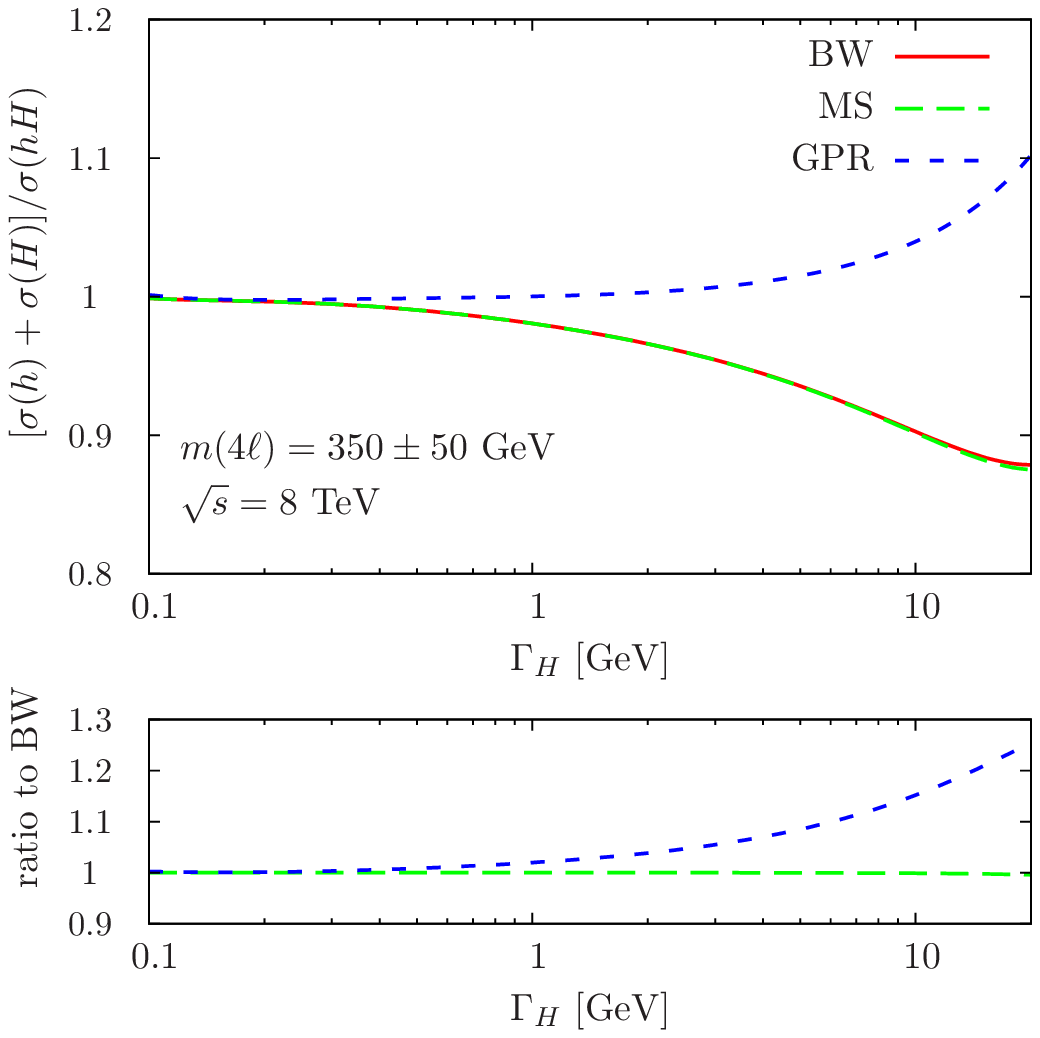}
    \caption{\label{fig:siginterf} Signal-signal ($h+H$) interference
      as a function of the total heavy Higgs decay width for the two
      parameter choices detailed in the text.}
    \end{center}
\end{figure*}
%%%%%%%%%%%%%%%%%%%%%%%%%%%%%%%%%%%%%%%%%%%%%%%%%%

The shape of the four lepton invariant mass distribution is mainly
driven by the particular choice of the Breit Wigner propagator in
Eq.~\eqref{eq:breitw}. Since this choice is ad-hoc, the
phenomenological implications do not have a theoretically well-defined
interpretation, especially when the interference with the $gg\to ZZ$
continuum is neglected~\cite{Hwidth2}. This is worsened by the fact
that we typically have a high precision for the
``signal''\footnote{The higher order QCD corrections to $H$ production
  directly generalize from the SM.} that is combined with comparably
lower precision for the ``background''.

There are suggestions to ameliorate this shortcoming by changing the
formulation of propagator for the signal
contribution~\cite{Hwidth,Hwidth2}, and we analyze these prescriptions
for two parameter choices
\begin{subequations}
  \begin{alignat}{5}
    \cos^2\chi=0.9 & \quad & m_H=180~\text{GeV}\,,\\
    \cos^2\chi=0.9 & \quad & m_H=350~\text{GeV}\,.
  \end{alignat}
\end{subequations}
in addition to the overall impact of interference. These choices are
motivated from current signal strength measurements~\cite{hzz} as well
as consistency with electroweak precision measurements
\cite{Baak:2014ora}, which prefer a small mixing and a rather light
state $H$. Furthermore, the mass choices coincide with the $Z$ boson
and top quark thresholds of the $gg\to h$ subamplitude, which make
these mass ranges particularly interesting due to an increase of the
continuum ({\it cf.}  Fig.~\ref{fig:mzz}).

To reflect finite detector acceptance\footnote{We perform an analysis
  in the fully leptonic final states but our findings are directly
  relevant for the boosted semi-hadronic analysis
  \cite{Hackstein:2010wk}.}, we cut on the four lepton invariant mass
to isolate the interference effects in this particular on-shell phase
space region
\begin{equation}
  m(4\ell)=m_H\pm 50~\text{GeV} 
\end{equation}
and choose three different approaches to include the width in our
calculation:
\begin{cedescription}
\item[Breit-Wigner (BW) propagator] Most calculations using
  multi-purpose Monte Carlo tools employ a Breit-Wigner propagator; we
  will use Eq.~\eqref{eq:breitw} as a reference.
\item[GPR prescription] A clean separation of signal and background
  has been proposed in Ref.~\cite{Hwidth2} by Gori, Passarino and
  Rosco. It is based on splitting the amplitude into a resonant and
  non-resonant part of the $2 \to 2$ scattering amplitude $pp\to ZZ$
    \begin{equation}
      A(s)=S(s) + B(s,t)
    \end{equation}
    with the ``signal'' defined as
    \begin{equation}
      S=V_{\text{prod}}(s_H) \Delta_H V_{\text{dec}}(s_H)\,.
    \end{equation}
    In this equation $s_H$ refers to the complex mass pole of the
    Higgs boson, {\it i.e.} the production and decay parts of the
    amplitude are evaluated at complex invariant masses, $s,t$ are the
    familiar Mandelstam variables and the propagator is then given by
    \begin{equation}
      \label{eq:newp}
      \Delta_H^{-1}(s)=s-s_H\,.
    \end{equation}
    As argued in Ref.~\cite{Hwidth2}, this prescription allows a
    theoretically robust matching of pseudo-observables between
    theory and experiment, and we refer the reader to this original
    publication for details. 

    Given the leading order nature of our calculation, there is a
    choice in defining $s_H$ which impacts the final result. We adopt
    the so-called ``bar'' convention (in particular to facilitate a
    comparison with the MS implementation below)
    \begin{equation}
      s_H= {m_H^2 - i m_H\Gamma_H \over 1+ \Gamma_H^2 / m_H^2}\,.
    \end{equation}
    An additional comment is necessary here because we will identify
    $m_H$ and $\Gamma_H$ with their on-shell parameters. The
    ``goodness'' of this identification is given by the ratio
    $\Gamma_H/m_H$: if the width becomes comparable to the mass, the
    bar scheme will deviate from the on-shell scheme. Since we are
    working in a tree-level setting, this choice is formally correct
    but higher order corrections are likely to quantitatively change our
    results when $\Gamma_H/m_H$ becomes large. In the following, we
    limit ourselves to parameters $\Gamma_H/m_H\lesssim 0.25$.
  \item[MS prescription] Seymour showed in \cite{Hwidth} that a simple
    modification of the propagator using a running width
    \begin{equation}
      \label{eq:ms}
      {1\over s - m_i^2} \to \left(1+i{\Gamma_i\over m_i}\right)
      \left( s-m_i^2+i{\Gamma_H\over m_i}s \right)^{-1}
    \end{equation}
    serves to reflect all relevant electroweak contributions in the
    high energy limit. In fact, this prescription is similar to the
    GPR implementation: Rewriting the propagator $s-s_H$ of
    Eq.~\eqref{eq:newp} using the definition of Eq.~\eqref{eq:newp}
    lets Eq.~\eqref{eq:ms} emerge in the bar-scheme. Note, however,
    that the substitution of Eq.~\eqref{eq:ms} does not imply an
    analytical continuation of production and decay subamplitudes to
    complex masses.
\end{cedescription}

Since we consider the fully leptonic final state (we neglecting QED
contributions), it should be noted that the $Z$ boson decay suffers
from similar shortcomings as discussed above \cite{Stuart:1991xk}. We
have explicitly checked the phenomenological impact of the analytic
continuation in the complex mass scheme and find a completely
negligible effect on the $pp\to 4\ell$ phenomenology and employ a
naive Breit-Wigner distribution for this part of the amplitude
throughout to allow for a consistent Higgs-specific comparison.

\section{Signal-Signal Interference}
\label{sec:hhinterf}
Let us first turn to ``signal-signal'' interference, {\it i.e.} the
interference between the two Higgs
bosons~\cite{bsm2,Maina:2015ela,kauernew}, of which the light SM state
$m_h=125~\text{GeV}$ acts as background. It should be noted that such
an analysis without including the $gg\to ZZ$ continuum is
incomplete~\cite{bsm,Hwidth2}, although in practical analyses as
performed by ATLAS and CMS such a discrimination is implicit.

In Fig.~\ref{fig:siginterf} we show the relative deviation
$[\sigma(h)+\sigma(H)]/\sigma(hH)$, which is directly sensitive to the
discussed interference. It can be seen that in the $ZZ$ threshold
region the interference effect can become of the order of $30\%$, and
depends crucially on the $h$ signal distribution as can be seen from
comparing the two parameter choices in Fig.~\ref{fig:siginterf}. 

The different treatment of the on-shell region in the discussed width
schemes induces a ${\cal{O}}(20\%)$ deviation as a function of the $H$
width for light states $m_H\lesssim 350~\text{GeV}$. The small
relative deviation of the BW and the MS scheme is directly related to
selecting a phase space region $s\sim m_H^2$, which induces a
modification $\sim \Gamma_H^2/M_H^2$ into the comparison. This ratio
is sufficiently small to not have a significant impact of the $H$
on-shell region for the considered parameter range. The main
difference of the GPR scheme in comparison to the other schemes is a
quantitatively changed behavior for $s\sim m_h$. The larger
$\Gamma_H$, the bigger this relative difference, a point already
stressed in the SM analysis of \cite{Hwidth2}.

\section{Signal-Signal-Background Interference}
\label{sec:allinf}
A crucial question, given the results of the previous section, is in
how far does the overall sensitivity to interference and the scheme
dependence of the previous section translate into a modification of
the total cross section when all interference effects are included?

%%%%%%%%%%%%%%%%%%%%%%%%%%%%%%%%%%%%%%%%%%%%%%%%%%
\begin{figure*}[!t]
  \begin{center}
    \includegraphics[width=0.43\textwidth]{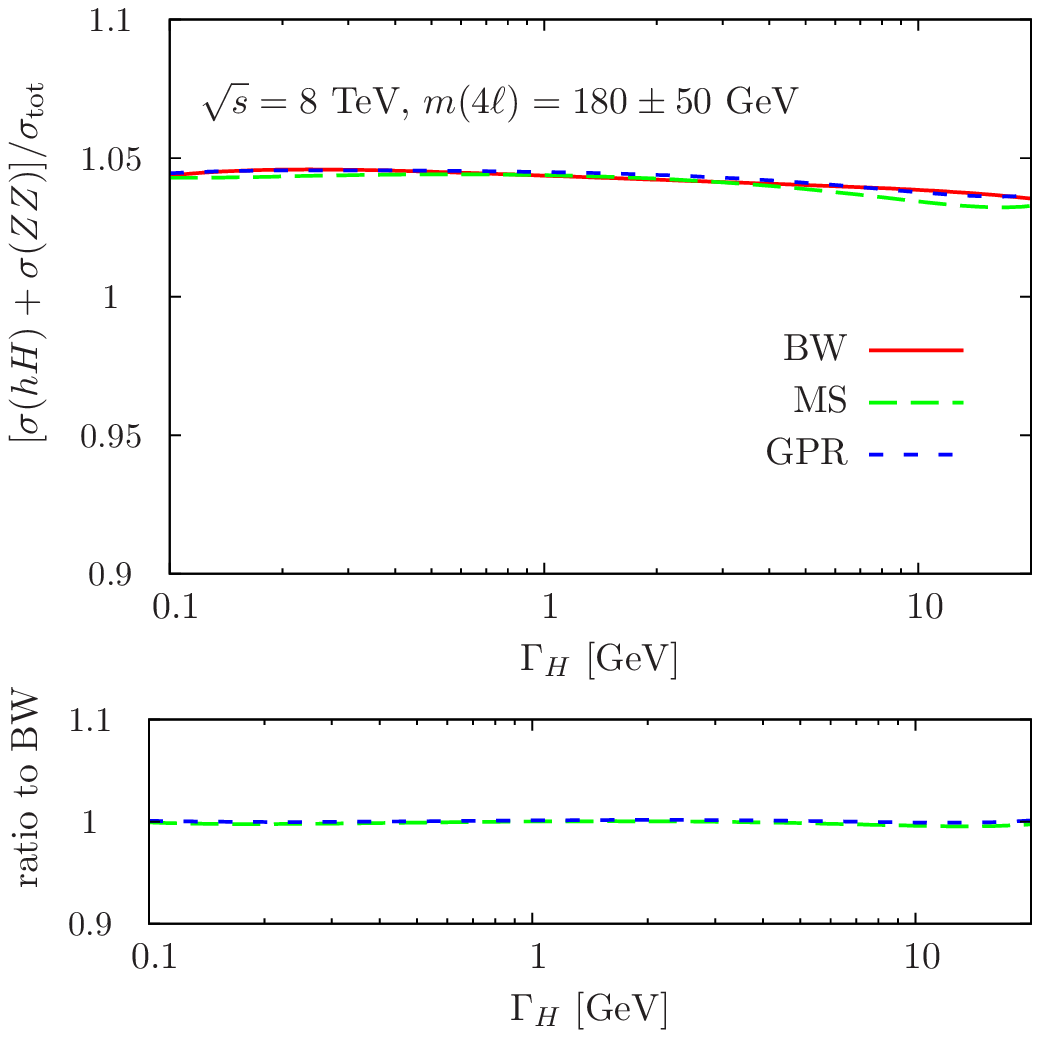}\hspace{1cm}
    \includegraphics[width=0.43\textwidth]{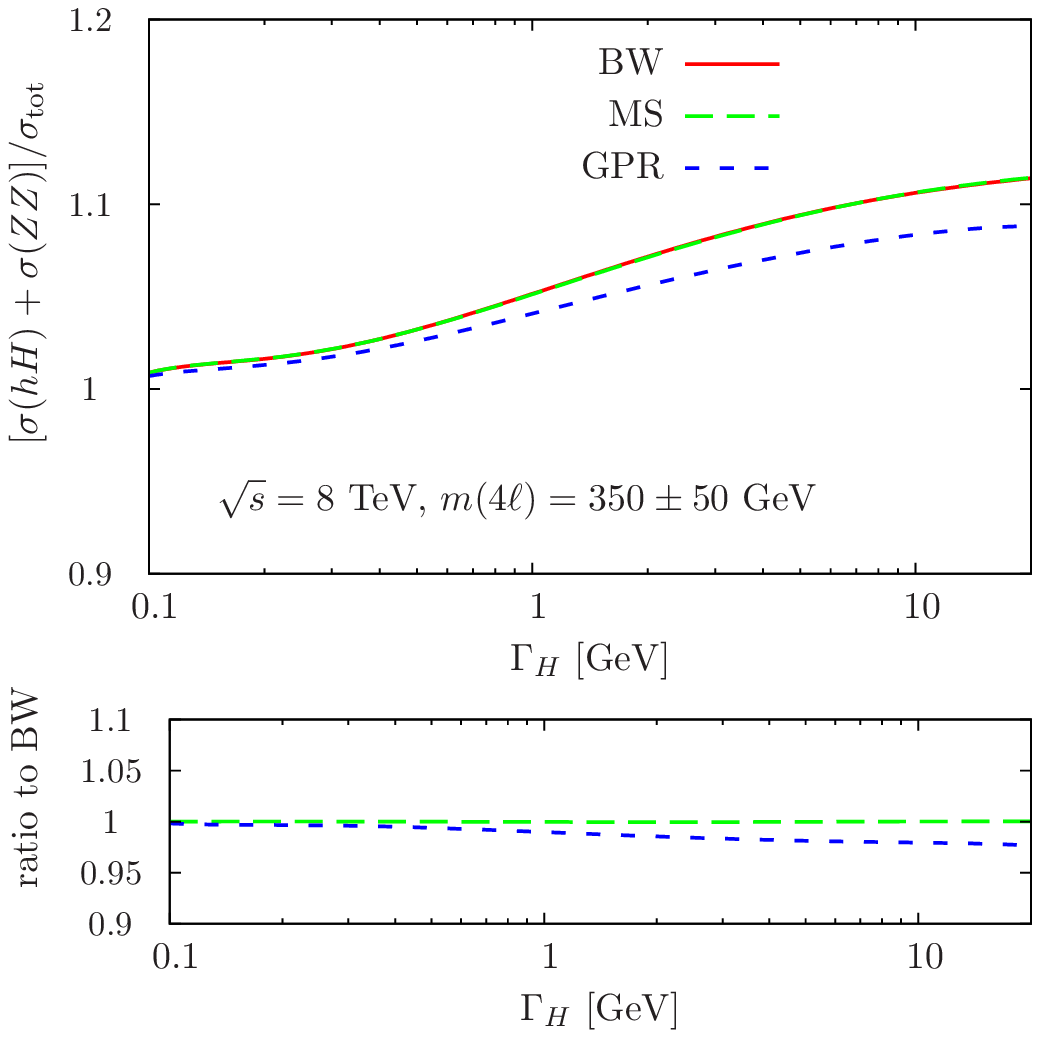}
    \caption{\label{fig:interf} $(h+H)$-continuum interference for the
      discussed prescriptions in the $H$ on-shell region, defined by
      the selection criteria on the final state invariant mass as
      shown.}
    \end{center}
\end{figure*}
%%%%%%%%%%%%%%%%%%%%%%%%%%%%%%%%%%%%%%%%%%%%%%%%%%

%%%%%%%%%%%%%%%%%%%%%%%%%%%%%%%%%%%%%%%%%%%%%%%%%%
\begin{figure}[!t]
  \begin{center}
    \includegraphics[width=0.43\textwidth]{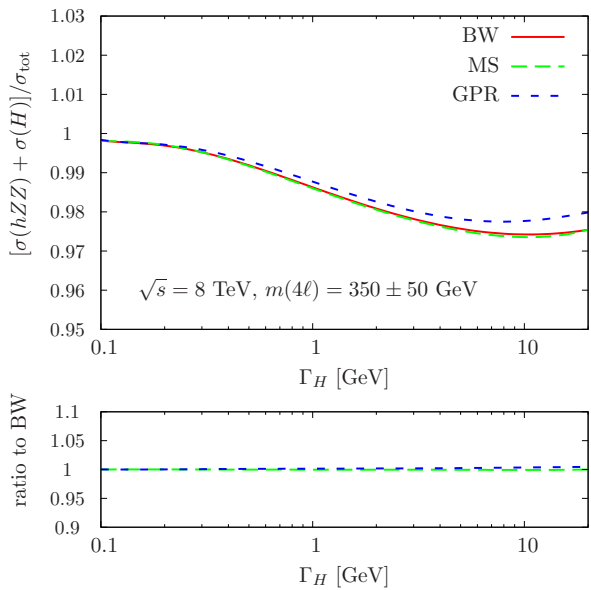}
    \caption{\label{fig:interf2} $(h+\text{continuum})-H$ interference
      for the discussed prescriptions in the $H$ on-shell region for
      the heavy state $m_H=350~\text{GeV}$.}
    \end{center}
\end{figure}
%%%%%%%%%%%%%%%%%%%%%%%%%%%%%%%%%%%%%%%%%%%%%%%%%%

On the one hand, interference of signal and background in $gg\to ZZ$
is known to be a sizable effect at large invariant final state masses
\cite{Kauer:2013qba,ciaran,bsm2}, ultimately as a sign of unitarity
and gauge invariance of the full scattering amplitude.\footnote{It
  should be noted that for inconsistent independent rescalings of
  gauge- and Yukawa sector Higgs couplings, the Lagrangian becomes
  ill-defined at scales as low as a few hundred GeV~\cite{bsm}.}
Hence, when integrating out the off-shell region, interference is
non-negligible~\cite{melnikov,cmswidth,atlaswidth}. On the other hand,
when considering the on-shell region at relatively moderate invariant
masses in a consistent electroweak model with only small deformations
compared to the SM phenomenology, the individual contribution of the
continuum can easily be 2 orders of magnitude above the signal
contribution before cancellations in the tail $pp\to ZZ$ above the
$t\bar t$ threshold become apparent (see
Refs.~\cite{Kauer:2013qba,ciaran}). As a consequence, the
modifications detailed in the previous section will be significantly
diluted if we consider the full final state. This is demonstrated in
Figs.~\ref{fig:interf} and \ref{fig:interf2}, which show the impact of
$hH-$continuum interference and the relative impact of the schemes
when we inject an $H$ signal to the $h$-continuum hypothesis for the
$m_H=350~\text{GeV}$ choice.\footnote{The difference for the
  $m_h=180~\text{GeV}$ spectrum is at the 1\% level due to the large
  continuum contribution.} The $\sim 30\%$ interference-induced
modifications reduce to an overall $\lesssim 10\%$ level with a scheme
dependence in the percent range. The former finding is consistent with
the results of \cite{kauernew} in support of the earlier claim of
\cite{bsm2} that on-shell interference is phenomenologically
subleading in high resolution channels at small $\Gamma_H/m_H$.

\bigskip

What is the phenomenological lesson to learn and how can experimental
results be impacted by our findings? Firstly, our parameter choices
are bound to a particular choice of mass scheme, which can only be
justified for relatively light $H$ masses that we discuss in this note
at the given (leading order) accuracy. Secondly, from a practitioner's
perspective, the overall impact of the interference effects are
tightly related to the treatment of systematic uncertainty treatment
in the actual analyses \cite{cmswidth,atlaswidth}. Currently, ATLAS
and CMS rely on leading order-precision in modelling the shapes of the
$gg\to ZZ$ distribution and the associated systematic uncertainty that
feeds into the limit setting are of the order of $25\%$. Even when we
rescale the individual signal and background contributions by total
$K$ factors as performed in Refs.~\cite{cmswidth,atlaswidth}, this
uncertainty is considerably bigger than the scheme and interference
dependence for our parameter choices. Hence, we can expect that the
current results should remain largely unaffected, but for analyses
with larger luminosities during run 2, interference effects should be
included.

We stress again that for heavy and wide $H$ candidates in the TeV
range the situation is qualitatively different. While such parameter
choices will automatically imply a tension with observed signal
strengths and electroweak precision data as soon as the signal
production cross section becomes large in the portal scenario, a
thorough inclusion of higher order corrections and a precise
definition of pseudo-observables following Ref.~\cite{Hwidth2} is
mandatory; a first step in this direction was presented in
Ref.~\cite{david}.

\section{Conclusions}
\label{sec:summary}

The search for new resonant contributions in the TeV regime is one of
the primary task of the LHC during the imminent run 2. Higgs
production with subsequent decay to leptons is one of the most
promising channels to facilitate a discovery of such a state in the
near future, with semi-hadronic $ZZ$ decays becoming an option for
larger $m_H$ values. Depending on the resolution and the width of such
an additional particle, additional interference effects and scheme
dependencies of this state should be included to consistently model
signal strengths and formulate exclusion limits, and to correctly
interpret a potential discovery.

%%%%%%%%%%%%%%%%%%%%%%%%%%%%%%%%%%%%%%%%%%%%%%%%%%
\acknowledgments
%\medskip
%\noindent {\it{Acknowledgements:}} 
We thank Nikolas Kauer and Claire O'Brien for discussions related to
their publication \cite{kauernew}.

C.E. is supported by the Institute for Particle Physics Phenomenology
Associateship program. I.L. is supported in part by the
U.S. Department of Energy under Contracts No. DE-AC02-06CH11357 and
No. DE-SC0010143. M.S. is supported in part by the European Commission
through the HiggsTools Initial Training Network PITN-GA-2012-316704.

%%%%%%%%%%%%%%%%%%%%%%%%%%%%%%%%%%%%%%%%%%%%%%%%%%
%%%%%%%%%%%%%%%%%%%%%%%%%%%%%%%%%%%%%%%%%%%%%%%%%%
%%%%%%%%%%%%%%%%%%%%%%%%%%%%%%%%%%%%%%%%%%%%%%%%%%

\end{document}